\documentclass[10pt,twocolumn]{article} 
\usepackage{times}
\usepackage{graphicx}
\usepackage{amssymb}
\usepackage{amsmath}
\usepackage{amsthm}
\usepackage{xcolor}
\newtheorem{thm}{Theorem}
\begin{document}

\title{A measure of compression gain for new alphabet symbols in data-compression}

\author{Richard  Fredlund\\
\\
\\
Richard\_Fredlund@hotmail.com\\
}

\maketitle
\thispagestyle{empty}

\begin{abstract}
Huffman encoding is often improved by using block codes, for example a 3-block would be an alphabet consisting of each possible combination of three characters. We take the approach of starting with a base alphabet and expanding it to include frequently occurring aggregates of symbols.  We prove that the change in compressed message length by the introduction of a new aggregate symbol can be expressed as the difference of two entropies, dependent only on the probabilities and length of the introduced symbol. The expression is independent of the probability of all other symbols in the alphabet. This measure of information gain, for a new symbol, can be applied in data compression methods. We also demonstrate that aggregate symbol alphabets, as opposed to mutually exclusive alphabets have the potential to provide good levels of compression, with a simple experiment. Finally, compression gain as defined in this paper may also be useful for feature selection.
\end{abstract}

\section{Introduction}
Shannon's theorem \cite{refi} states that the minimum encoding for a given alphabet of symbol codes is the Shannon entropy times the number of symbols to be encoded.  However, simply taking the character frequencies in a text document (the message to be encoded) and applying Huffman encoding \cite{refh} or Arithmetic \cite{refg, refj} encoding to it, which would closely approximate the Shannon minimum encoding for the given alphabet (of individual characters) does not achieve best overall compression for the message. This is because natural languages (and many other data types) contain interdependencies, which are not modelled simply by the symbol frequencies.  In cases where one symbol informs another, as in natural language, the fully conditional entropy is the lower bound of compressibility \cite{refi}.  

For an alphabet to encode a message it must be complete in the sense that all fragments of the message can be described by it. For this reason n-block Huffman encoding is sometimes used as a way to boost the compression of a direct Huffman encoding. For example a $3$-block would be an alphabet consisting of each possible combination of three characters. It is known that as the block lengths increase the encoding converges towards the lower bound of compressibility \cite{refi}. However as the block length grows, the number of symbols in the alphabet grows exponentially becoming practically infeasible for block lengths bigger than five. Also large N-block alphabets would contain many redundancies in the form of character sequences which may never appear in the language.

The symbol alphabet used for most data compression schemes are either individual characters, syllables (for example \cite{refa,refb}), or words (for example \cite{refc, refd}). With a few technical exceptions (e.g. punctuation in word encodings) these alphabets essentially form a partition of the text into more or less, similar parts.  Spaces have been added to words \cite{refe} and frequently occurring $2$,$3$ and $4$ grams have also been used to pad out the alphabet, \cite{reff}. The class of all variable-length-to-variable-length finite state encoders are considered in \cite{refk} to demonstrate that the universal algorithm of Lem-Ziv approaches asymptotically the lower bound of compressibility. Lem-Ziv extracts sequences of variable length however these are not encoded directly by the Huffman (or Arithmetic) encoder. Rather, the pointers to sequence location and sequence length are encoded.

In contrast we use direct Huffman encoding with variable length aggregate symbols (DH-VLAS) to encode the message. Starting with a base alphabet, aggregate symbols of arbitrary length are added to the alphabet for encoding. The potential advantages of this approach are\footnote{There is also a hypothetical motivation behind this approach; which is the possibility of creating generic alphabets for specific data domains allowing very rapid on-the-fly compression and decompression while maintaining reasonable rates of compression.}:
\begin{enumerate}

\item To create alphabets with good rates of direct compression without the feasibility issues of large N-block Huffman encoding, 
\item Fast decompression and therefore easy access to compressed data; which may have applications in data-base storage, where access times matter but space is at a premium.
\item The method is universal to any data type with repetitions.
\item In contrast to Lem-Ziv and other algorithms the essential structure of the original message is preserved under compression.
\end{enumerate}

We start by taking a base alphabet and expanding it to include frequently occurring aggregates of base symbols. When an aggregate is added the minimum encoding for the alphabet given by the Shannon entropy times the number of symbols to be encoded will change. It may increase or decrease depending on the symbol added. This is because the number of symbols in the message will decrease (as more than one symbol is combined into just one new aggregate symbol, for every occurrence in the message) but the entropy of the alphabet will increase. For example when encoding simple text, composed of $26$ alphabet symbols, at what frequency is the compound symbol 'the' composed of three alphabet symbols worth adding to the model? To realise that for high enough frequencies adding the symbol may produce a decrease in the expected message length notice that the number of 'symbols' in the message will decrease by two times the number of occurrences of the word 'the'. As each occurrence of the word 'the' represents a condensing of three alphabet symbols into just one new symbol. If the frequency of 'the' is high enough in the message then the decrease in message length will outweigh the increase in entropy and the new alphabet will have a lower minimum encoding. 

We define the compression gain for a given aggregate symbol as the change in minimum encoding associated with the addition of a new aggregate symbol. The compression gain for a new aggregate symbol may be positive or negative depending on it's frequency in the message. The disadvantage of this formulation for compression gain is that it depends on the frequency of all characters in the alphabet. 

The main contribution of this paper is to find a simple formulation for the compression gain, which is concise and independent of the frequency of all other characters from the alphabet not in the new symbol. Computation time for this measure is therefore independent of size of the alphabet.  This is shown Section $2$ where we prove that the average gain, as measured in bits per character, of introducing a new symbol S can be expressed as the difference of two entropies plus a correction term which involves only the probability and length of S; where each of the entropies is expressed only in terms of the relative frequencies of the alphabet symbols which are being combined to form S. 

In Section $3$, we perform an experiment to find aggregate symbols for the compression of Alice.txt, from the Canterbury Corpus. The purpose of this experiment is to demonstrate the use of the information gain measure to create an aggregate alphabet for data compression; and also to show that direct Huffman encoding using variable length aggregate symbols (DH-VLAS) has the potential to produce good rates of compression. The experiment is not optimal, and so does not represent a lower bound for DH-VLAS.  However, starting with a base alphabet of just $72$ distinct characters and an alphabet optimal compression rate of $4.5$ bits per character. We discover an alphabet with $1319$ aggregate symbols which effectively reduces the rate of compression to under $3$ bits per character.

\section{Main Result}
Let us start by considering a message $M=m_1,m_2,...,m_N$ composed of $N$ characters from a base alphabet $A=\{a_1,a_2,...,a_K\}$ with known character frequencies $F=\{f_{a_1},f_{a_2},...,f_{a_K} \}$. The probabilities of each character symbol are $P = \{ p_{a_1},p_{a_2},...,p_{a_K}\}$ where $p_{a_i}=f_{a_i}/N$ for each $i \in \{1,2,...,K\}$.
By Shannon's information theory \cite{refi} the minimum encoding length for the message with this alphabet is: 
\begin{equation}
\label{eqA}
L(M,A)= N \sum_{x \in A} p_x \log_2 (1/p_x) := N * H(A) 
\end{equation}
where $H(A)$ is the entropy of the message $M$ for this probability model, described by $\{A,P\}$.   

We now propose the addition of a new symbol to the alphabet. Let $S = b_1 b_2 ... b_r$ be an aggregate symbol composed of $r$ symbols from the original alphabet $A$ so that $b_i \in A$ for $i \in \{1,2,...,r\}$, (Note that $b_i$ and $b_j$ may not be distinct), and let $f_s$ be the number of non-overlapping occurrences of $S$ in the message $M$, and $\{S\}$ be the set of distinct characters in $S$.

Adding $S$ to the alphabet gives a new alphabet of character symbols $A' = \{S,a_1,a_2,...,a_K\}$, a new set of probabilities $P' =\{p'_s,p'_{a_1},...,p'_{a_K}\}$ and $F' = \{f'_S,f'_{a_1},...,f'_{a_K}\}$.By counting the number of characters $a_i$ in the original message and subtracting occurrences which appear in $S$ we get the updated character frequencies:

\begin{equation}
\label{eqB}
f'_{a_i} = f_{a_i} - f_S S(a_i)
\end{equation}
and probabilities
\begin{equation}
\label{eqC} 
p'_{a_i}=p_{a_i}-p_S S(a_i)
\end{equation}
where $S(a_i)$ is the number of times the character symbol $a_i$ appears in $S$. Note for characters not in $S$ the number of occurrences will remain unchanged and $f'_{a_i}=f_{a_i}$.
The message length $N$ will also change. This is because, for each of the $f_S$ occurrences of $S$ in the message, $r$ symbols from the original character sequence $M$ will be condensed into just one character in the new alphabet $A'$ This gives a new message length of: 
\begin{equation}
\label{eqD}
N' = N - f_S (r-1)
\end{equation}
Alternatively, letting $\mu_s = (1-p_s(r-1))$ the length of the new message $N'$ is $\mu_s N$ and we call $\mu_s$ the rescale constant.

Therefore, the minimum encoding length for the new alphabet, containing the aggregate symbol $S$, will be:
\begin{equation}
\label{eqE}
L(M,A')= N' \sum_{x \in A'} p'_x \log_2 (1/p'_x) := N' * H(A'). 
\end{equation}
Notice that L(M,A') may be more or less than L(M,A) depending on the frequency of $S$. We define the compression gain associate with the introduction of symbol $S$ to be:
\begin{equation}
\label{eqF}
\text{gain}(S) = L(M,A)-L(M,A'),
\end{equation}
and the compression gain per character $\Delta(S)$ to be:
\begin{equation}
\label{eqG}
\Delta(S) = \frac{\text{gain}(S)}{N}.
\end{equation}
Notice that $\Delta(S)$ will be positive if the addition of $S$ to the alphabet results in a compression gain, and negative otherwise. Notice that in its current form the expression for $\Delta(S)$ depends on the frequency of all characters in the alphabet, regardless of whether they are contained in $S$ or not. 

We therefore prove the following concise theorem which shows that $\Delta(S)$ can be expressed as the difference of two entropies, which depend only on the probabilities of symbols in $S$ and the probability of $S$. In other words it is independent of the probabilities of all other symbols in the alphabet:

\begin{thm}
The compression gain $\Delta(S)$ associated with adding a symbol for $S$ is given: 
\begin{equation}
\label{eqH}
\begin{split}
 \Delta(S) &= H(\mu_S ,p_{a_i} : a_i \in \{S\})\\
&-H(p_S,p'_{a_i} \mu_S :a_i \in \{S\})
\end{split}
\end{equation}

\end{thm}

\begin{proof}
We start by noting equation \ref{eqE} can be re-written in terms of the character frequencies: 
\begin{equation}
\label{eqI}
L(M,A')= N' \log_2 N - \sum_{x \in A'} f'_x \log_2 f'_x
\end{equation}
using the identity $\log_2 (1/x) = -\log_2 (x)$ and by noting that the sum of the frequencies $\sum_{x \in A} f_x$ is equal to N. This can be further split by grouping characters in $S$ and those not in $S$ separately:

\begin{equation}
\label{eqJ}
\begin{split}
  L(M,A') &= N' log_2 N' - f_S  \log_2 f_S   
\\&-\sum_{a_i \in \{S\}} f_{a_i} \log_2 f_{a_i}
\\&+ \sum_{a_i \in \{S\}} f'_{a_i} \log_2 f'_{a_i}
\end{split}
\end{equation}
We have split the summation for the log frequencies into three distinct parts.  Firstly, the term containing $f_S$ the number of occurrences of the newly introduced character $S$. Secondly, those characters not in $S$ whose frequency do not change with the introduction of $S$ and finally, those characters which are in $S$ and whose frequencies do change. 

The advantage of this formulation is that the terms involving frequencies not in $S$ cancel and:

\begin{equation}
\label{eqK}
\begin{split}
\text{gain}(S) &= N \log_2 N - N' \log_2 N'  
 \\ &+ f_S \log_2 f_S
\\&-\sum_{a_i \in \{S\}} f_{a_i} \log_2 f_{a_i}
\\&+ \sum_{a_i \in \{S\}} f'_{a_i} \log_2 f'_{a_i}
\end{split}
\end{equation}
The disadvantage of this formulation is that the expected compression gain per symbol, $\Delta(S)$, appears to be dependent on $N$, due to the $\log_2 N$ term. To see that this ought not to be the case consider the scenario of a new message $M_{new}$ which is precisely $M$ repeated twice, such that all character frequencies are exactly doubled.  We would expect to see the overall gain for this new message to be $2 * \text{gain}(S) $. 

We now recast equation \ref{eqK} in terms of probabilities: this is achieved by replacing $f_s$ with $Np_s$,  $N'$ with $N\mu_S$, and $f'_{a_i}$ with $p'_{a_i}\mu_S N $. 

\begin{equation}
\label{eqL}
\begin{split}
  gain(S) &= N \log_2 N - \mu_S N \log_2 \mu_S N  
 \\ &+ p_S N \log_2 p_S N
\\&-\sum_{a_i \in \{S\}} p_{a_i}N \log_2 p_{a_i}N
\\&+ \sum_{a_i \in \{S\}} p'_{a_i}\mu_S N \log_2 p'_{a_i}\mu_S N
\end{split}
\end{equation}

This simplifies by noticing that all terms are of the form $AN \log_2 AN$ and using the identity $AN log_2 AN = AN log_2 N - AN\log_2 1/A$. Summing up the $N\log_2 N$ terms everything cancels by noting that $p_S * r = \sum_{a_i \in \{S\}}p_S S(a_i)$.  This gives a sum over terms all of the form $A\log2 A$, the first of which is $1 \log_2 1 = 0$ which also cancels to give: 

\begin{equation}
\begin{split}
\Delta(S) &= \mu_S \log_2 (1/\mu_S) - p_S  \log_2 (1/p_S)  
\\&+\sum_{a_i \in \{S\}} p_{a_i} \log_2 (1/p_{a_i})
\\&- \sum_{a_i \in \{S\}} p'_{a_i}\mu_S \log_2 (1/p'_{a_i}\mu_S)
\end{split}
\end{equation}  
In more concise notation this can be re-written as:
\begin{equation}
\begin{split}
\Delta(S)  &= H(\mu_S ,p_{a_i} : a_i \in \{S\})\\
&-H(p_S,p'_{a_i} \mu_S :a_i \in \{S\})
\end{split}
\end{equation}
Which completes the proof.
\end{proof}

\section{ Alphabet encoding scheme}

We start with a very basic standard alphabet of just $64$ characters, which are assumed to be known previously by both the encoding and decoding software. This character set is simply the uppercase and the lowercase letters, and $12$ punctuation symbols.  These are, space, newline, colon, semi-colon, exclamation mark, open and close brackets, hyphen, comma, full-stop and question-mark. 

The first $5$ bits of the alphabet encoding are used to encode, $MXBITS$, the number of bits required to encode a frequency count in this document.  The maximum being $32$ bits assigned to each frequency count, which would allow up to $4294$ million occurrences of any character within the document.  $MXBITS$ is determined by the maximum number of occurrences of any character in the document.  For Alice.txt, $MXBITS$ is set to $15$ indicating that $15$ bits should be used to record the frequency of each alphabet symbol.  Note that in practice this is only the maximum ever required, and in practice most symbol frequencies for this document fall into the range $0$-$255$ and could be expressed using only $8$ bits).

Next each of the $64$ characters from the standard alphabet is assigned $15$ bits to record it's frequency in the document (960 bits). 

The next $8$ bits in the encoding are assigned to $NUMCHAR$, the number of non-standard characters in the document.  We keep things simple here and assume all other characters in the document are $8$ bit ascii codes (unicode symbols are not considered).

In Alice.txt there are $8$ characters not included in the standard alphabet.  These are apostrophe, two forms of speech-mark, '*', open and close square brackets, and the numerals, 2 and 9. 

These are encoded with $8$ bits for ascii code, and $15$ bits for character frequency.  That is 23 * 8 = 184 bits. 

Before any aggregate symbols are added therefore, the number of bits used by the alphabet encoding is $5+960+8+184=1157$.
The next $15$ bits of the encoding are allocated to $AGCOUNT$, the number of aggregate symbols to be included in the alphabet. 

At initialization, $AGCOUNT$ is set to $0$ and the alphabet contains $64+8=72$ characters.  Aggregate symbols in this implementation consist of two ordered symbols from the already existing alphabet.  As new aggregates are added they may consist of combinations of previous aggregates, or of characters from the base alphabet.  As the aggregate alphabet increases in size, the number of possible aggregates also increases.  For an alphabet of length $N$ the first and second symbol both have $N$ possible values giving a total of $N^2$ possible new aggregate symbols.  The number of bits required to encode each aggregate symbol depends on the number of bits required to encode a value in $\{1,...nN\}$. At initialization $N=72$, hence the number of bits required to encode each sub-symbol in the aggregate is given by $SUBCOUNT$ equal to $\log_2 (AGCOUNT +72)$ rounded up to the nearest integer. $15$ bits must also be allocated to record the frequency count.  This gives a total of $15 + 2 * SUBCOUNT$ bits for each new aggregate symbol.  This starts off at $29$ bits and increases by $2$ bits each time $N$ gets too large. 

\section{Search algorithm}

Here we have kept the search for aggregate symbols with maximum compression gain as simple as possible.  At each step the symbol gain is calculated for each of the possible $N^2$ aggregates and the highest gain is selected.  If this gain is higher than the number of bits necessary for the alphabet encoding then the symbol is added. 

This approach becomes infeasible as $N$ grows large, and it is expected that better search algorithms can be found to find aggregates with a high expected gain. 

\section{Results}
The brute force algorithm described above was implemented on Alice.txt from the Canterbury Corpus.  The uncompressed document contains $148480$, $8$ bit ascii characters.  At initialisation there were $72$ distinct characters in the base alphabet.  The alphabet encoding occupied $1157$ bits and the optimal message encoding length which could be closely approximated by arithmetic encoding was $670057$ bits.  So the total message encoding before any aggregates were introduced was $671214$ bits.  This gives a message encoding of $4.51$ (2.d.p.) bits per character. Subsequently $1319$ aggregate symbols were added to the alphabet.  This increased the number of bits required by the alphabet encoding to $46674$ bits.  However, the optimal message encoding dropped to $390937$ bits, leaving a total encoding length of $427511$ bits.  This gives a message encoding of $2.63$ (2.d.p) nits per character and a total encoding of $2.95$ (2.d.p.) bits per character. 

The bit rate could have dropped further however by this point the experiment had slowed down considerably, due to the brute force search, so the experiment was stopped soon after the bit rate of $3$ bits per character was reached. 

The above table shows the first $20$ aggregate symbols selected by the algorithm. Notice that the first $18$ symbols seem quite generic and might well be informative symbols for many text documents.  It is interesting to note that the second symbol is 'the' while 'th' is not selected until the 18th symbol; instead 'he' is selected as the first symbol, having high frequency and an information gain of $6538$ bits. 

However, symbol $19$ and $20$ are 'Al' and 'Ali' and the next two symbols are 'Alic' and 'Alice' respectively.  These symbols are specific to the document.  Other aggregate symbols, which are recognisable as being specific to Alice.txt and appear in the alphabet are 'Dormouse', 'Hare', 'Mock', 'curious', 'remark', 'said Alice', 'little', 'the Queen' and 'Duchess'. 

Notice that the information gain is not always strictly decreasing with each new aggregate symbol.  The information gain of 'ing', 'and' and 'Ali' are each higher than their immediate predecessors, 'in', 'nd' and 'Al', respectively.  Meaning that where a longer aggregate symbol has high gain, the shorter aggregates which build up to the longer symbol may have a lower gain. 
\begin{center}
 \begin{tabular}{|c|c|c|c|} 
 \hline
  n & gain & symbol & freq \\ 
 \hline\
 1 & 6538 & he & 3705 \\ 
 2 & 4694 & the & 2101 \\
 3 & 3249 & in & 1998 \\
 4 & 4051 & ing & 979 \\
 5 & 3098 & ou & 1515 \\  
 6 & 3054 & [sp] the & 1938 \\ 
 7 & 2135 & , [sp] & 1761 \\ 
 8 & 2020 & [nl][nl] & 827 \\ 
 9 & 2167 & . [nl][nl] & 404 \\ 
 10 & 2064 & ve & 663 \\ 
 11 & 1968 & nd & 1172 \\ 
 12 & 2648 & and & 880 \\ 
 13 & 1899 & ,' & 397 \\ 
 14 &  1995 & ' [nl][nl] & 343 \\ 
 15 & 1792 & , [sp] and & 418 \\ 
 16 & 1693& [sp] - & 262 \\ 
 17 & 1553 & [sp]the[sp] & 1314 \\ 
 18 & 1392 & th & 1096 \\ 
 19 & 1345 & Al & 403 \\ 
 20 & 1832 & Ali & 395 \\ 
 \hline
\end{tabular}
\end{center}

 \section{Discussion}
The aggregate alphabet found, achieves a $41.7$\% (1.d.p.) reduction in message encoding length at, and a $34.7$\% (1.d.p.) reduction in total encoding length relative to the Shannon minimum encoding for the base alphabet of single characters.  Reaching a message encoding of $2.63$ (2.d.p.) bits per character and a total encoding of $2.95$ (2.d.p.) bits per character before the experiment was stopped. 

Both the alphabet encoding and the search algorithm used to find new aggregate symbols are not ideal and leave considerable scope to be improved upon, in developing a practical data compression scheme. However the results obtained demonstrate the basic efficacy of using this measure of information gain to create an alphabet for data compression. 

The two drawbacks of this implementation are: firstly, it can only find larger chunks which are informative by working up through smaller chunks first. e.g. 'Al','Ali','Alic','Alice'. (which requires the smaller chunks to also be informative); secondly, once a larger chunk is found the smaller chunks which lead up to it will often become less informative in the alphabet. However it is interesting that the pairwise-encoder is able to find some longer sequences such as 'the Queen', and 'said Alice', showing that there is enough information gain in the pairwise steps to lead up to these particular sequences. 

Better compression rates could be achieved by using a dictionary approach and initially adding whole words before starting the pairwise combination of symbols, and also removing smaller chunks if they become uninformative. 

\section{Conclusion}
The mathematical contribution in Section $2$ has been to find a simple formulation which measures the change in compression length when adding a new aggregate symbol to an alphabet code for date compression.  This formulation is concise and independent of the frequency of all other characters from the alphabet not in the new symbol.  Computation time for this measure is therefore independent of size of the alphabet. 

This measure is potentially applicable to any variable length coding scheme with an alphabet and could be used with either static or adaptive schemes such as PPM \cite{refj}.

It is also worth noting that compression gain as defined here can also be viewed as a measure of 'informativeness' of a symbol sequence, based on both frequency and length. The words 'Alice', 'Dormouse', 'Hare', 'Mock', 'curious', 'remark', which have high compression gain also seem informative of the document content.  It is therefore likely a good measure for feature selection, both in text classification \cite{refl} and in general.

Finally, while text-compression has been used as the specific example here the true flexibility of this approach is that it can be a applied to any data-type containing redundancies.

\bibliographystyle{abbrv}
\bibliography{asme2e}
\end{document}